\documentclass[superscriptaddress,aip,amsmath,amssymb,floatfix,reprint,raggedbottom]{revtex4-1}
\usepackage{graphicx}
\usepackage[colorlinks=true,citecolor=blue,linkcolor=blue,urlcolor=blue]{hyperref}
\usepackage{amsmath}
\usepackage{dcolumn}
\usepackage{xcolor}
\usepackage{fancyhdr}
\usepackage{lipsum}
\usepackage{soul}
\usepackage{balance}
\usepackage{footnote}

\makeatletter 
\renewcommand{\fnum@figure}{\textbf{Fig.~\thefigure}}
\makeatother

\makeatletter
\def\bbordermatrix#1{\begingroup \m@th
  \@tempdima 4.75\p@
  \setbox\z@\vbox{%
    \def\cr{\crcr\noalign{\kern2\p@\global\let\cr\endline}}%
    \ialign{$##$\hfil\kern2\p@\kern\@tempdima&\thinspace\hfil$##$\hfil
      &&\quad\hfil$##$\hfil\crcr
      \omit\strut\hfil\crcr\noalign{\kern-\baselineskip}%
      #1\crcr\omit\strut\cr}}%
  \setbox\tw@\vbox{\unvcopy\z@\global\setbox\@ne\lastbox}%
  \setbox\tw@\hbox{\unhbox\@ne\unskip\global\setbox\@ne\lastbox}%
  \setbox\tw@\hbox{$\kern\wd\@ne\kern-\@tempdima\left[\kern-\wd\@ne
    \global\setbox\@ne\vbox{\box\@ne\kern2\p@}%
    \vcenter{\kern-\ht\@ne\unvbox\z@\kern-\baselineskip}\,\right]$}%
  \null\;\vbox{\kern\ht\@ne\box\tw@}\endgroup}
\makeatother

\begin{document}
\title{ Implementing Bayesian networks 
with embedded stochastic MRAM}
\author{Rafatul Faria}
\thanks{Equally contributing authors}
\affiliation{School of Electrical and Computer Engineering, Purdue University, West Lafayette, Indiana 47907, USA}
\author{Kerem Y.  Camsari}
\thanks{Equally contributing authors}
\affiliation{School of Electrical and Computer Engineering, Purdue University, West Lafayette, Indiana 47907, USA}
\author{Supriyo Datta}
\email{datta@purdue.edu}      
\affiliation{School of Electrical and Computer Engineering, Purdue University, West Lafayette, Indiana 47907, USA}
\date{\today}

\begin{abstract}
Magnetic tunnel junctions (MTJ's) with low barrier magnets have been used to implement random number generators (RNG's) and it has recently been shown that such an MTJ connected to the drain of a conventional transistor provides a three-terminal tunable RNG or a  $p$-bit. In this letter we show how this $p$-bit can be used to build a $p$-circuit that emulates a Bayesian network (BN), such that the correlations in real world variables can be obtained from electrical measurements on the corresponding circuit nodes. The $p$-circuit design proceeds in two steps: the BN is first translated into a behavioral model, called Probabilistic Spin Logic (PSL), defined by dimensionless biasing (h) and interconnection (J) coefficients, which are then translated into electronic circuit elements. As a benchmark example, we mimic a family tree of three generations and show that the genetic relatedness calculated from a SPICE-compatible circuit simulator matches well-known results.
\end{abstract}
\maketitle


Magnetic tunnel junctions (MTJ's) with low barrier magnets have been used to implement random number generators (RNG's) \cite{fukushima2014spin,choi2014magnetic,lee2017design,parks2017superparamagnetic} and it has recently been shown that such an MTJ connected to the drain of a conventional transistor provides a three-terminal tunable RNG or a  \textit{p-bit}  \cite{camsari2017implementing} with applications to optimization \cite{sutton2017intrinsic} and an enhanced type of Boolean logic, that is invertible \cite{faria2017low,pervaiz_emulations,pervaiz2017weighted,camsari2017stochastic}. In this paper we show how this $p$-bit can be used to build a $p$-circuit that emulates a Bayesian network (BN)  \cite{pearl2014probabilistic} defined in terms of conditional probability tables (CPT) that describe how each \textit{child node} is influenced by its   \textit{parent nodes}. BN's are widely used to understand causal relationships in real world problems such as forecasting, diagnosis, automated vision, manufacturing control and so on \cite{heckerman1995real}. For deep and complicated networks where each child node has many parent nodes, the computation of the joint probability becomes impractical \cite{heckerman1996causal} and different hardware implementations of BN's have been proposed \cite{rish2005adaptive,chakrapani2007probabilistic,zermani2015fpga,tylman2016real,weijia2007pcmos,shim2017stochastic, friedman2016bayesian, querlioz2015bioinspired, behin2016building}. 

In this letter we present a systematic approach for translating a BN into an electronic circuit such that the stochastic node voltages mimic the real world variables whose correlations can be obtained from electrical measurements on the corresponding circuit nodes. The proposed electronic circuit and the hardware building blocks are based on present day Magnetoresistive Random Access Memory (MRAM) technology whose MTJs are built out of thermally unstable nanomagnets \cite{camsari2017implementing} (Stochastic MRAM), obviating the need for the development of a new  device. 
\begin{figure}
\centering
\includegraphics[width=0.95\linewidth]{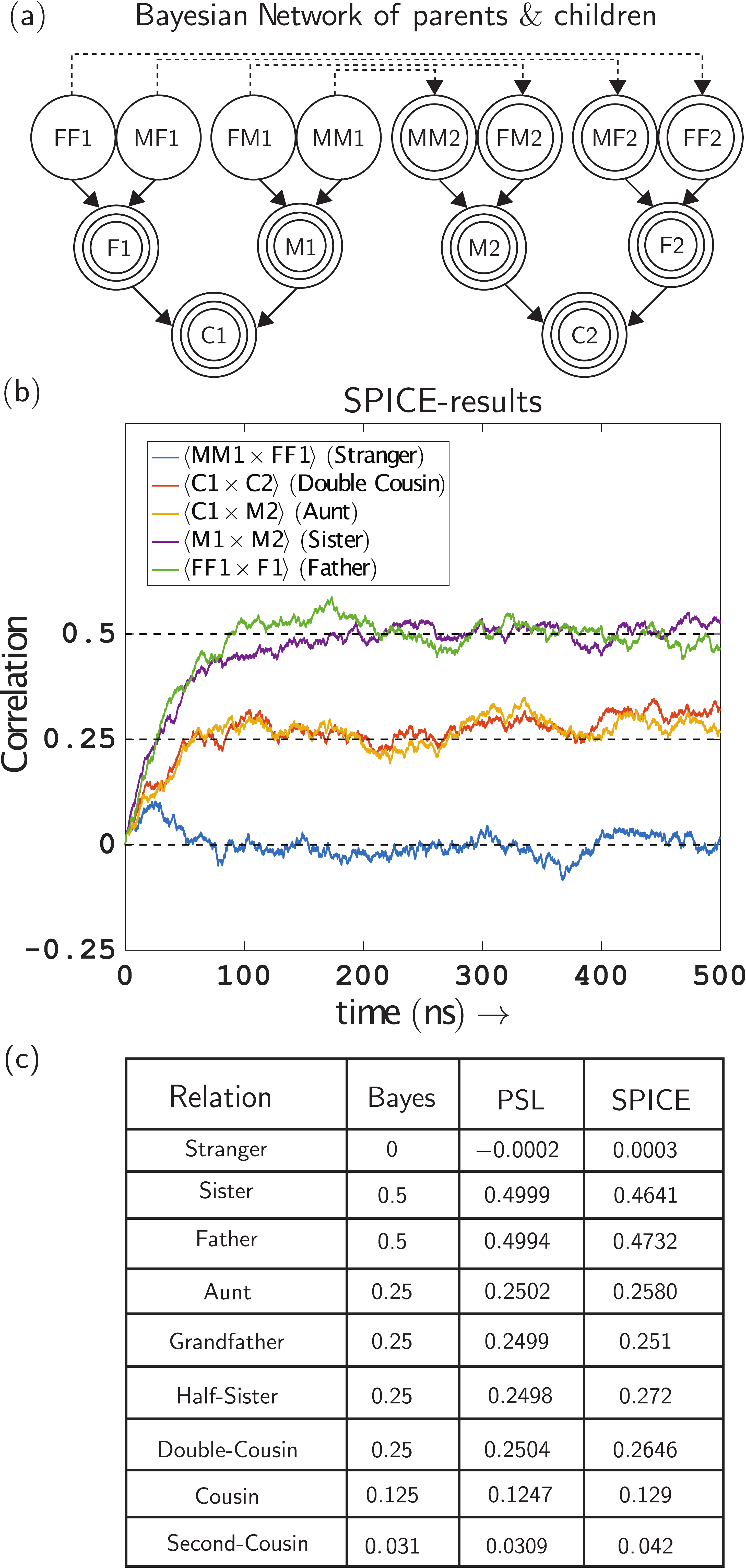}
\caption{\textbf{(a) An example Bayesian Network (BN)} showing three generations with children, parents and grandparents. The grandparent generation has no explicit parents, but we can introduce their correlations implicitly by making the second set of grandparents (MM2,FM2,MF2,FF2) conditionally dependent on the first set (FF1,MF1,FM1,MM1) as shown. The rest of the nodes (C1, F1, M1, C2, F2, M2) are each conditionally dependent on two parents. \textbf{(b) Representative SPICE-results} from the full hardware circuit of Fig.~\ref{fi:fig4} when the circuit is set up so that FF1 $\approx$ FF2,  MF1 $\approx$ MF2, FM1 $\approx$ FM2,  MM1 $\approx$ MM2. In this scenario, $C1$ and $C2$ are double cousins. \textbf{(c) Relatedness of family members} calculated from three different models: a behavioral model, PSL, a SPICE model for the corresponding circuit and the well-known result from standard statistical arguments applied to BN. Single, double and triple encirclements indicate a zero-parent node, one parent node, and two parent node respectively, as indicated in Fig.~\ref{fig2}.}
\label{fig1}
\end{figure}

As a benchmark example, consider a BN (Fig.~\ref{fig1}) consisting of three generations of a family, where each child ($C$) inherits half the genes from the father ($F$) and the other half from the mother (M), so that $C=0.5F + 0.5M$, where $C$, $F$ and $M$ can each be viewed as a bipolar random variable: ($-$1,$+$1). A well-known concept in genetics is that of  \textit{relatedness}. For example, the relatedness $\langle C1\times C2 \rangle$ of two siblings, with the same parents is $50\%:$ 
$\langle C1\times C2 \rangle = .25 \big( \langle F\times F \rangle+\langle F\times M \rangle+\langle M\times F \rangle+\langle M\times M \rangle \big)
= .25 \big(1+0+0+1 \big) = 0.5.$ On the other hand two cousins whose fathers are siblings have a relatedness of only $12.5\%$: $\langle C1\times C2 \rangle = .25 \big( \langle F1\times F2 \rangle+\langle F1\times M2 \rangle+\langle M1\times F2 \rangle+\langle M1\times M2 \rangle \big)
= .25 \big(0.5+0+0+0 \big) = 0.125.$ Fig.~\ref{fig1}b compares the well-known relatedness of different family members (see for example, Ref. \cite{nolan_web}) with that calculated from a behavioral model which we call probabilistic spin logic, PSL, and from a simulation of the actual circuit using a SPICE-based circuit simulator.

\begin{figure}[t!]
\includegraphics[width=0.99\linewidth]{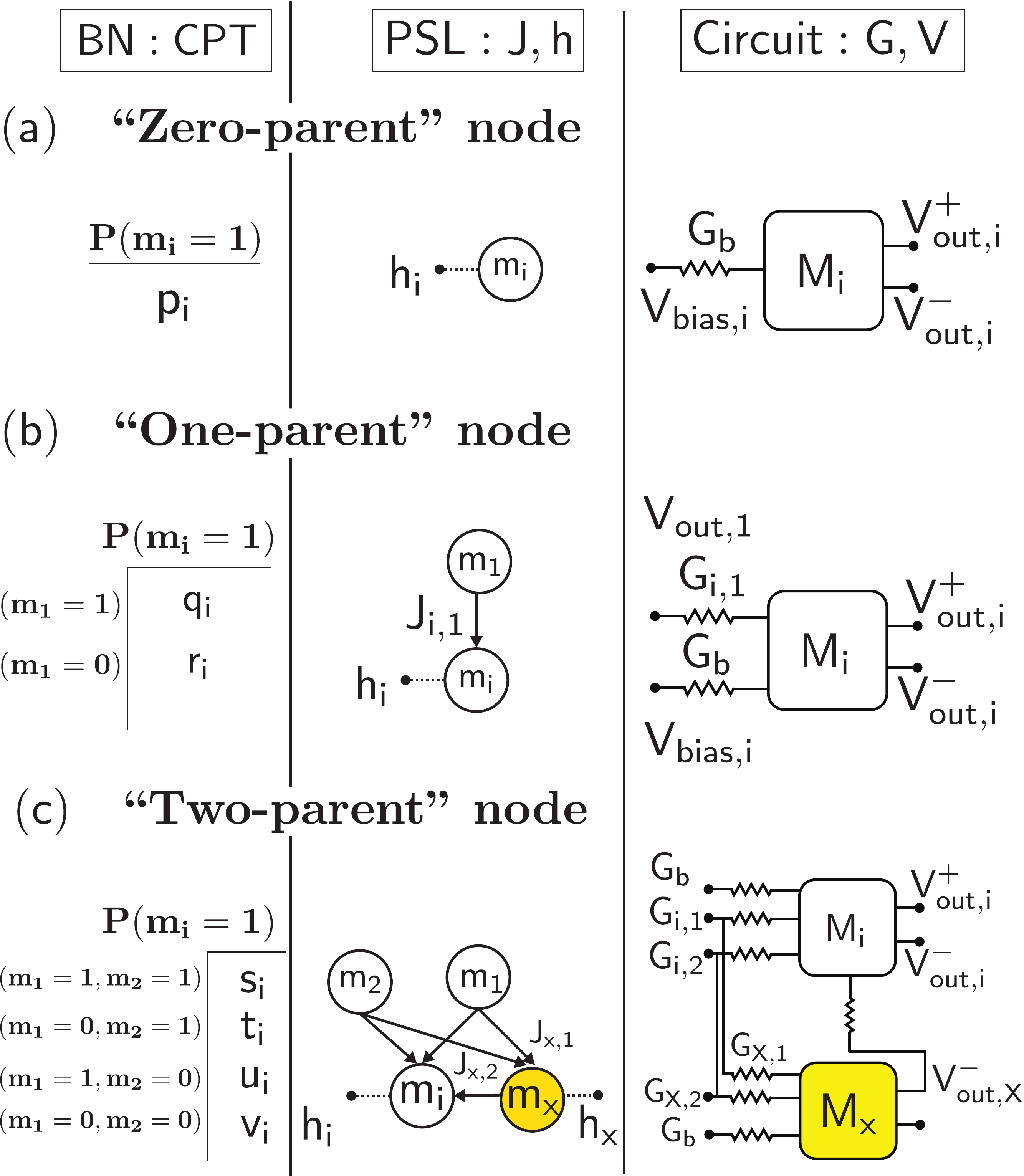}
\caption{\textbf{Translating nodal information from BN to PSL to circuit}:  Each node of a BN is described by a conditional probability table (CPT), that of a PSL network is described by dimensionless constants $J,h$, and that of circuit is described by conductances $G$ and voltage $V_{\rm bias}$. The text describes how the CPT is translated to $J,h$ and then to $G,V_{\rm bias}$ for (a) zero-parent node, (b) one-parent node and (c) two-parent node.}
\label{fig2}
\end{figure}

The behavioral PSL model represents an intermediate step in the translation of BN's to electronic circuits. It is a network whose nodes are abstract $p$-bits denoted by $m$ (see Fig.~\ref{fig2}) connected to other nodes and biased through dimensionless constants $J, h$ respectively. The $p$-bits described by Eq.~\ref{eq:PSL}a is analogous to a binary stochastic neuron and their interconnection described by Eq.~\ref{eq:PSL}b is analogous to a synapse. The PSL model is then translated into a circuit model whose nodes are actual circuit elements denoted by $M$ connected to other nodes and biased through conductances $G$ and voltages $V_{\rm bias}$. Fig.~\ref{fig1}b shows that the relatedness from the PSL model (second column) as well as that obtained from the SPICE model (third column) agree well with the standard BN result (first column), thus providing confidence that the circuits obtained following our procedure can be used to study BN's in general.

Genetic relatedness is a textbook concept that provides a good benchmark for a hardware circuit emulator, but the principles presented here can be used to emulate more complicated BN's as well, involving more complex CPT's, as well as more complex nodes with $N>2$ parents, reflecting the presence of more than two factors influencing the occurrence of an event.

\section{Probabilistic Spin Logic: Behavioral Model}
PSL is defined by two equations  \cite{camsari2017stochastic} loosely analogous to a neuron and a synapse. The former  is a binary stochastic neuron, or what we call a $p$-bit, whose output $m_i$ is related to its dimensionless input $I_i$ by the relation
\begin{subequations}
\begin{equation}
m_i (t+\Delta t) = \mathrm{sgn} \big( \mathrm{rand}(-1,1) + \mathrm{tanh} \ I_i (t) \big) 
\end{equation}

\noindent where $\mathrm{rand} \ (-1,+1)$ is a random number uniformly distributed between $-$1 and +1, and $t$ is the normalized time unit. The synapse generates the input $I_i$ from a weighted sum of the states of other $p$-bits according to the relation
\begin{equation}
{I_i}(t) = I_{0} \bigg( h_i(t)+\sum_j{J_{ij}m_j} \bigg)
\end{equation}
\label{eq:PSL}
\end{subequations}
\noindent where, $h_i$ is the on-site bias and $J_{ij}$ is the weight of the coupling from $j^{th}$ $p$-bit to $i^{th}$ $p$-bit.
\section{From BN nodes to PSL nodes}
To relate $I_i$ to the conditional probability $P_i$ for $m_i$ to be 1, we note from Eq.~\ref{eq:PSL}a that the average value of $m_i$ is $\mathrm{tanh}(I_i)$ and this must equal $P_i \times (+1) + (1-P_i) \times (-1)$=$2P_i-1$. Making use of Eq.~\ref{eq:PSL}b we can write
\begin{equation}
 I_0 (h_i+\sum_{j} J_{ij}m_j)=\mathrm{tanh}^{-1} (2P_i-1) 
\label{p_J}
\end{equation}
\noindent We use this relation to translate the $P_i$ from the CPT into $J, h$ in the PSL model, but the details differ depending on the number of ``parents'' of node $i$ (Fig.~\ref{fig2}).

Nodes with \textit{no parents} have no connecting weights $J_{ij}$, only a bias $h_i$ which is related to the specified conditional probability $p_i$ by $h_i = (1/I_0) \ \mathrm{tanh}^{-1} (2p_i-1)$. Nodes with \textit{one parent} have one connecting weight $J_{ij}$, and a bias $h_i$ which can be obtained from the two specified conditional probabilities $q_i,r_i$ from the equations
\begin{subequations}
\begin{equation}
h_i+J_{i1}(+1)= \frac{1}{I_0} \mathrm{tanh}^{-1}(2q_i-1)
\end{equation}
\begin{equation}
h_i+J_{i1}(-1)= \frac{1}{I_0} \mathrm{tanh}^{-1}(2r_i-1)
\end{equation}
\end{subequations}
\normalsize
\noindent Nodes with \textit{two parents} have two connecting weights $J_{i1},J_{i2}$, and a bias $h_i$ but there are four equations for these three unknowns. All equations can be satisfied simultaneously only if the equations are not linearly independent. If they are independent then an auxiliary node $X$ is introduced so that:
\footnotesize
\begin{subequations}
\begin{equation}
h_i+J_{i1}(+1)+J_{i2}(+1)+J_{iX} m_{X} = \frac{1}{I_0} \mathrm{tanh}^{-1}(2s_i-1)
\end{equation}
\begin{equation}
        h_i+J_{i1}(-1)+J_{i2}(+1)+J_{iX} m_{X} = \frac{1}{I_0} \mathrm{tanh}^{-1}(2t_i-1)
\end{equation}
\begin{equation}
h_i+J_{i1}(+1)+J_{i2}(-1)+J_{iX} m_{X} = \frac{1}{I_0} \mathrm{tanh}^{-1}(2u_i-1)
\end{equation}
\begin{equation}
h_i+J_{i1}(-1)+J_{i2}(-1)+J_{iX} m_{X}=t= \frac{1}{I_0} \mathrm{tanh}^{-1}(2v_i-1)
\end{equation}
\label{eq:J}
\end{subequations}
\normalsize

\noindent where $m_{X}=\mathrm{tanh}(h_X+J_{X1} m_{1}+J_{X2} m_{2})$ with the parents $m_{1},m_{2}$ equal to ($\pm1,\pm1$) as appropriate for the four equations. One possibility is to choose $h_{X},J_{X1},J_{X2}$ such that $m_{X}= m_{1} \cap m_{2}$ and then select the four remaining unknowns $h_{i},J_{i1},J_{i2},J_{iX}$ to satisfy Eqs.~\ref{eq:J}.

Nodes with $N$ parents have a total of $(N+1)$ unknowns, but there are $2^N$ equations to satisfy. Depending on the number of linearly independent equations, it is necessary to introduce the appropriate number of auxiliary variables.  In this letter we will only present results for the BN in Fig.~\ref{fig1} which includes nodes with a maximum of $N=2$ parents. Moreover, the CPT for the 2-parent nodes is assumed to be of the form $t=u=0.5$, $s=1-\varepsilon$ and $v=\varepsilon$, $\varepsilon$ being a small number introduced to avoid the singularities associated with the $\mathrm{tanh}$ function. With this CPT, no auxiliary node (X) is needed.
\section{From PSL nodes to circuit nodes}
\begin{figure}[t!]
\includegraphics[width=0.95\linewidth]{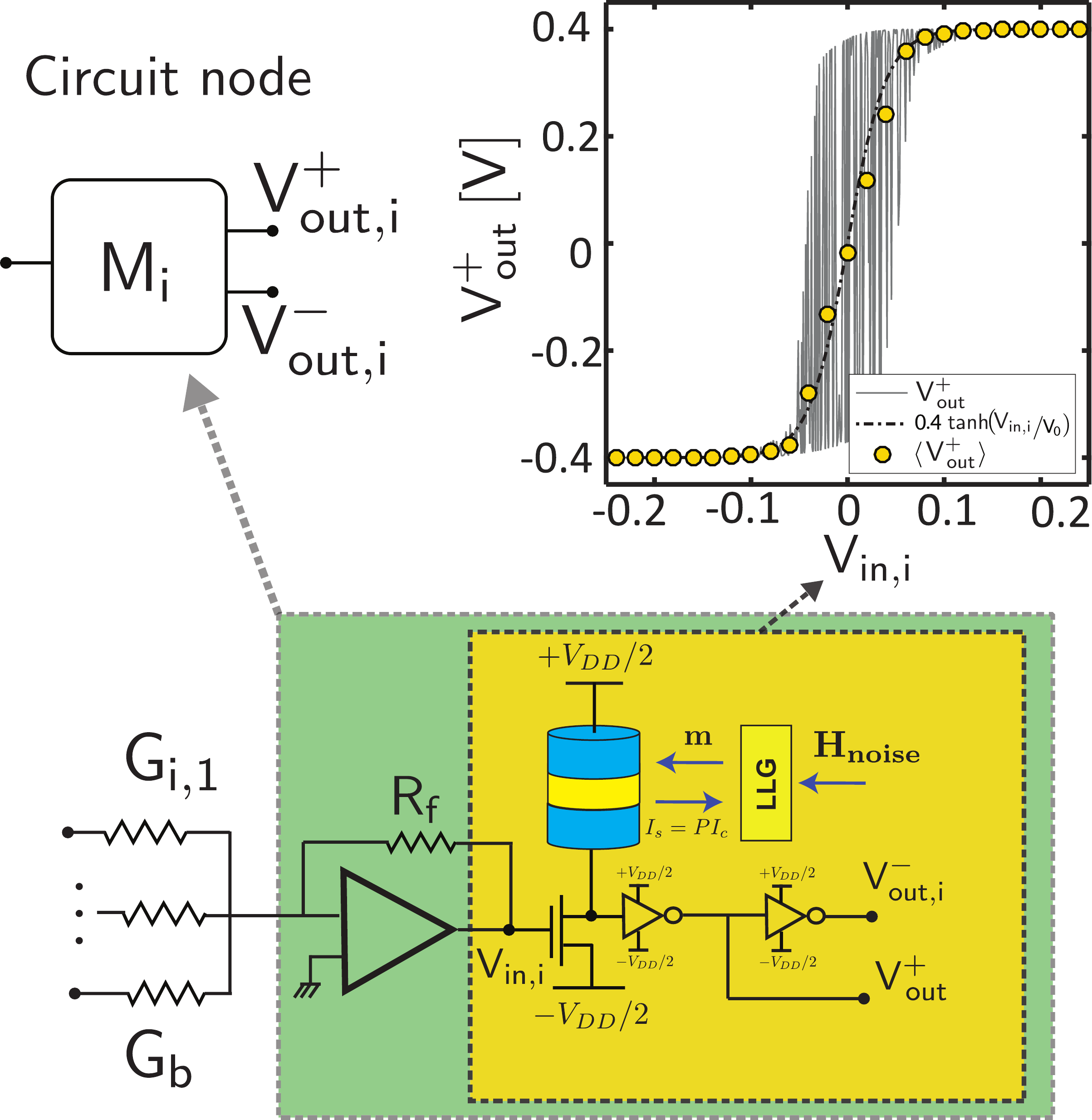}
\caption{\textbf{Circuit implementation of building block}:The circuit Eqs.~\ref{eq:circuit} can be mapped onto the PSL Eqs.~\ref{eq:PSL} using Eqs.~\ref{eq:map} as described in the text. The circuit node $M_i$ is defined to include the transimpedance amplifier along with the $p$-bit. The details of the embedded MRAM based p-bit are discussed in the text.}
\label{fig3}
\end{figure}
\begin{figure}[t!]
\centering
\includegraphics[width=0.99\linewidth]{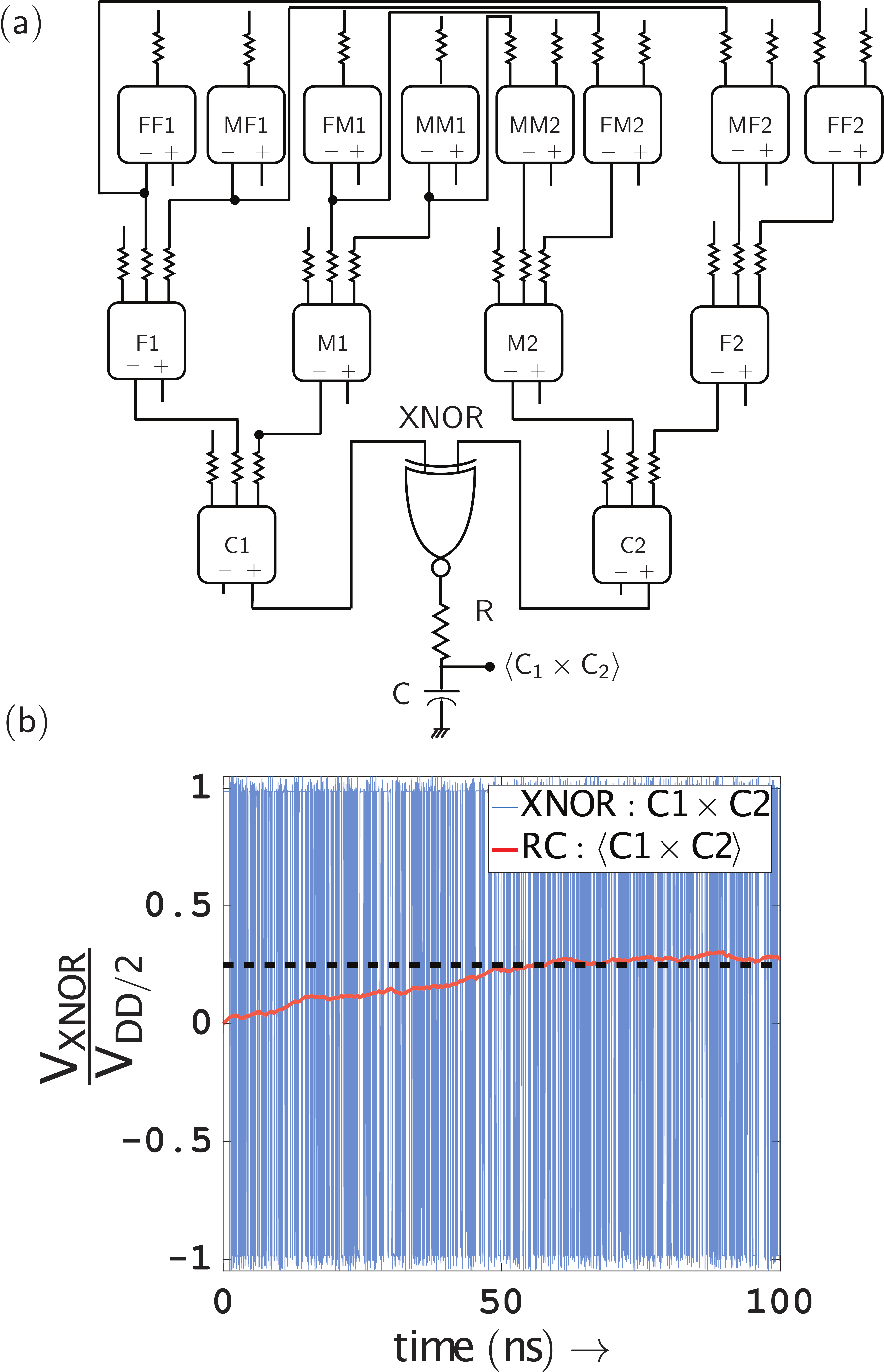}
\caption{\textbf{SPICE simulation of the full circuit} designed to mimic the Bayesian network in Fig.~\ref{fig1}a. (a) Circuit diagram, (b) Typical stochastic nodal voltages from which nodal correlations can be obtained using an XNOR gate and a long time constant RC circuit. In the present example, the following parameters are used: The RC circuit uses $R=200\rm \ k \Omega$, C=200 fF, $R_f=150 \rm \  k\Omega$ and $I_0=1$ with dimensionless weights $J_{ij}=J_0= 2.3026$ which are then used to obtain conductances $G_{ij}$ from Eq.~\ref{eq:map}. A simulation time of 1 ps is used in HSPICE that combines the self-consistent stochastic LLG with Predictive Technology models (PTM) \cite{predictive_tech} as in \cite{camsari2017implementing}. All transistors use the 14nm HP-FinFET node with minimum fin numbers (nfin=1). The XNOR gate is designed as a standard 14 transistor CMOS circuit, inverting an XOR output.}
\label{fi:fig4}
\end{figure}
To translate the PSL into a circuit we use the embedded MTJ  \cite{camsari2017stochastic} whose output is related to its input by the relation
\begin{subequations}
\begin{equation}
V_{out,i}= \frac{V_{DD}}{2}  \mathrm{sgn} \left( \mathrm{rand}(-1,+1) + \mathrm{tanh} \frac{V_{in,i}}{V_{0}} \right) 
\label{eq:pbit}
\end{equation}
\noindent where $\pm V_{DD}/2$ are the supply voltages, and $V_0$ is a parameter ($\sim 50 \rm \ mV$) describing the width of the sigmoidal response. Although $V_0$ is a fitting parameter for the tanh function, it captures the actual sigmoidal response of the MTJ unit quite well. Even if there is a slight deviation with the tanh function due to the skewness of the MTJ response, it will not cause a noticeable difference in the output since PSL is quite robust against noise \cite{camsari2017stochastic}. The output voltages are connected back through conductances $G$ with a transimpedance amplifier having a feedback resistor $R_f$, so that (see Fig.~\ref{fig3})
\begin{equation}
V_{in,i} = V_{\rm bias,i} G_b R_f + \sum_j{V_{out,j} G_{ij} R_f}
\end{equation}
\label{eq:circuit}
\end{subequations}
\noindent Eqs.~\ref{eq:circuit} can be mapped onto the PSL Eqs.~\ref{eq:PSL} by defining
\begin{subequations}
\begin{equation}
m_i = \frac{V_{out,i}}{V_{DD}/2}, \ \  I_i = \frac{V_{in,i}}{V_0}
\end{equation}
\begin{equation}
h_i = \frac{V_{\rm bias,i}}{V_{DD}/2}, \ \ J_{ij} = \frac{G_{ij}}{G_{b}}, \ \ I_0 = G_b R_f \frac{V_{DD}} {2V_0}
\end{equation}
\label{eq:map}
\end{subequations}
\normalsize
\section{SPICE-based p-bit Model}
In order to design the basic building block for the BN based on Eq.~\ref{eq:pbit}, we are following the p-bit design in Ref.~\cite{camsari2017implementing} that describes an embedded MTJ structure with a stochastic nanomagnet. We consider the weight logic in Eq.~\ref{eq:PSL}b to be implemented using ideal transimpedance amplifier with resistors  \cite{camsari2017stochastic} though a capacitive network with a more compact implementation could also be used to implement the weighted sum operation \cite{hassan2018voltage}. We use the same parameters for the p-bit as in \cite{camsari2017implementing}: A circular (stochastic) in-plane nanomagnet with negligible uniaxial anisotropy ($H_K\sim0$) \cite{cowburn1999single,debashis2016experimental}, damping coefficient for the nanomagnet $\alpha=0.01$, saturation magnetization $\rm M_s=1100$ emu/cc, with a free layer diameter 22 nm and a thickness of 2 nm. A Tunneling Magnetoresistance (TMR) value of 110\% is used based on \cite{lin200945nm}. The MTJ conductance is assumed to be bias-independent and is given by $G(t)=G_0\big[1+m_z(t)\rm \ TMR/(2+TMR)\big]$, where $m_z(t)$ is provided to the model by a self-consistent solution of the sLLG (stochastic Landau-Lifshitz-Gilbert equation) solver. The device operation is based on the control of the transistor conductance through the input voltage. The non-linear transistor characteristics with respect to drain, gate and source voltages are captured in simulation by the 14 nm HP-FinFET node from the Predictive Technology Models (PTM) \cite{predictive_tech}. When the transistor conductance is much greater or less than the MTJ conductance, the output shows little noise but when the MTJ conductance is matched to the transistor ON resistance around $V_{in,i}$=0, there are large fluctuations at the output. In Fig.~\ref{fig3}, each circular dot in the sigmoid is obtained by averaging  1 $\mu$s  response of the stochastic output and the dashed lines show a (tanh) fit with a $V_0=50$ mV. The solid lines are obtained by sweeping the input voltage rail-to-rail in 100 ns and plotting the input with respect to the output voltage. Within the modular SPICE framework, the magnetization dynamics of the circular stochastic nanomagnet is captured by solving the sLLG equation in the macrospin assumption,
\begin{subequations}
\begin{align}
&(1+\alpha^2)\frac{d\hat m}{dt} = -|\gamma|{\hat m \times \vec{H}} - \alpha |\gamma| (\hat m \times \hat m \times \vec{H})\nonumber \\ &+  \frac{1}{q  N}(\hat m \times \vec{I}_{S} \times \hat m)  + \left(\frac{\alpha}{q N} (\hat m \times \vec{I}_{S})\right)
\label{sLLG}
\end{align}
\end{subequations}
\noindent where $\alpha$ is the damping coefficient, $\gamma$ is the electron gyromagnetic ratio, $N=\rm M_s$Vol./$\mu_B$ is the total number of Bohr magnetons in the magnet volume, $\rm M_s$ is the saturation magnetization, $\vec{H}=\vec{H_d}+\vec{H_n}$ is the effective field including the out-of-plane ($\hat x$ directed) demagnetization field $\vec{H_d}=-4\pi \rm M_s m_x \hat x$, as well as the thermally fluctuating magnetic field due to the three dimensional uncorrelated thermal noise $H_n$  with zero mean $\langle{H_n}\rangle=0$ and standard deviation $\langle H_n^2\rangle= 2\alpha \rm kT / |\gamma| M_sV$ along each direction, $I_S=PI_C\hat z$ is the spin current along the MTJ fixed layer direction ($\hat z$) where $P$ is the polarization of the fixed magnet. The model takes this spin-current ($I_S$) incident to the free layer into account and for the parameters we have used, this current does not cause appreciable pinning of the free layer. A time step $\Delta t=1$ ps is used for the circuit simulation which implies a noise bandwidth of $\Delta f=1$ THz.
\section{SPICE-based Circuit Model}
Fig.~\ref{fi:fig4}a shows the full circuit assembled using the nodes defined in Fig.~\ref{fig3} to mimic the Bayesian network in Fig.~\ref{fig1}a. Fig.~\ref{fi:fig4}b shows typical nodal voltages obtained from a SPICE simulation, whose correlations can either be calculated in software or measured using an XNOR gate to multiply them as shown and finding the long term average with an RC circuit having a time constant $\gg$ T:
\begin{equation*}
\langle C1\times C2 \rangle= \int_{0}^{T} \frac{dt}{T} \ C_1 (t) C_2 (t)
\end{equation*}
These nodal correlations in the circuit can be used to compute the correlation between causally connected real world variables. For example the relatedness of different family members cited in Fig.~\ref{fig1}b were all obtained in this manner from circuit simulations. Different relationships between $C1$ and $C2$ are enforced by using different CPT's for their grandparents. For example, if all grandparents are unrelated, the grandchildren $C1$ and $C2$  would show zero correlation. But if we enforce perfect correlation between $FF1$, $FF2$ and between $MF1$,$MF2$ through the corresponding CPT, we effectively make $F1$ and $F2$ into siblings with a correlation of $50\%$. $C1$ and $C2$ then are first cousins with a correlation of $12.5\%$.If we further enforce perfect correlation between $FM1$, $FM2$ and between $MM1$, $MM2$, we also make $M1$ and $M2$ into siblings with a correlation of $50\%$, just like $F1$ and $F2$.  $C1$ and $C2$ now are double cousins with a correlation of $25\%$. 

Note that this is an asynchronous circuit with no clocks of any kind. This is particularly interesting since the corresponding PSL simulations require $p$-bits to be updated sequentially from parent to child node. In the SPICE circuit simulation there is no central clock to enforce an updating sequence, but our results show that the correlations  are in good agreement with the PSL results and with Bayes theorem. However, such an asynchronous operation works only if the interconnect delays, for example from node FF1 to FF2, are much shorter than the nanomagnet fluctuations as discussed in Reference~\cite{pervaiz_emulations}. Since magnetic fluctuations occur at $\sim$ns time scales,  this condition is naturally satisfied. The slight mismatch of the Bayes theorem and  the PSL model appears to decrease systematically with increasing sample size (N=1e7 for the examples shown in Fig.~\ref{fig1}b) with the full circuit model closely following them, but the updating issue in asynchronous operation deserves further study. We have not considered variations in the thermal barriers or interconnect delays in this paper, which requires further study.
\section{Conclusions}
In summary, we have used SPICE simulations to show that using existing MRAM technology it should be possible to build \textit{p}-circuits that mimic Bayesian networks such that each stochastic node is represented by a stochastic \textit{p}-bit. We show that the \textit{ensemble-averaged}  correlations between the actual physical variables can be estimated from the \textit{time-averaged} correlations between the voltages at the corresponding nodes which are measured electronically with XNOR logic gates and long time constant RC circuits, thus requiring no software-based processing of any kind. Our results could open up a new application space for Embedded MRAM technology with minimal modifications.

\section{Acknowledgment}
This work was supported by the National Science Foundation (NSF).

\balance

 \end{document}